\documentclass[prb,aps,twocolumn,showkeys,showpacs]{revtex4}
\usepackage{epsf}
\usepackage{graphicx}
\usepackage{amsmath}

\begin{document}
\title{Integrated light in direct excitation and energy transfer luminescence}

\author{Eugeniusz Chimczak}

\affiliation{Pozna\'n University of Technology, Faculty of Technical Physics, 
ul. Nieszawska 13 a, 60-965 Pozna\'n, Poland}

\date{\today}
\email{chimczak@phys.put.poznan.pl}

\begin{abstract}
Integrated light in direct excitation and energy transfer luminescence
has been investigated. In the investigations reported here,
monomolecular centers were taken into account. It was found
that the integrated light is equal to the product of generation
rate and time of duration of excitation pulse for both direct
excitation and energy transfer luminescence.
\end{abstract}

\maketitle

\section{INTRODUCTION}
Spectral and kinetic measurements are used to determine the
characteristic features of the luminescence as well as the luminophors
investigated. However, contrary to the spectral measurements, the kinetic
measurements allows very often to give a quantitative  description of the
dependence investigated permitting the better determination of the
mechanism of luminescence and, for example, electroluminescent
device. Therefore, many papers are devoted to kinetic of
luminescence~\cite{chimczak84,bertrandt95,chimczak89,chimczak89acta,
berdowski85chem,wang03,kimpel95,roura94,bhargava94,krasnov01,john96,
barthou94,zhang95PRB,busse76,seo03,weman95,harukawa00,joubert87,lucas94,
wolfert85,berdowski85,suchocki87,leslie81,chimczak85,zhang95,bergman87,
lee92,ramirez05,aizawa06,kapoor00,thilsing05}. Some investigators
have established correlated formulas to describe the time dependence
of luminescence. For example, Chimczak \emph{et al.} (1984, 1995) described
theoretically very well time dependence of ZnS:Mn electroluminescence
of thin film devices excited by short voltage pulse~\cite{chimczak84,
bertrandt95}. The researches (1989) have also obtained good fitting
of the electroluminescence to the experimental data during~\cite{chimczak89}
as well as after the end~\cite{chimczak89acta} of rectangular exciting pulse.
Berdowski \emph{et al.} (1985) obtained very good description of emission of
$\textrm{Tb}^{3+}$ in $\textrm{CsCdBr}_{3}:\textrm{Tb}^{3+}$ upon
excitation in ${}^{5}D_{3}$ level at 1.3 and 75 K~\cite{berdowski85chem}.
Wang \emph{et al.} (2003) have derived an analytical expression
constructed as a sum of several products of Gauss function and exponential
functions to describe the time dependence in $\textrm{ZnS:Er}^{3+}$ thin
films~\cite{wang03}. Kimpel \emph{et al.} (1995) explained time behavior
of the $\textrm{Cr}^{2+}$ in ZnS~\cite{kimpel95}. Roura \emph{et al.}
have analyzed dynamics of the infrared photoluminescence in silicon
powder~\cite{roura94}. All the cited above investigations allows obtaining
integrated light of luminescence. 
However, up to now, nobody has described the integrated light, which
is additional kind
of investigations of luminescence kinetic. The aim of the paper is to describe
the integrated light for direct excitation and energy transfer luminescence
assuming monomolecular centers in the description.

\section{DIRECT EXCITATION LUMINESCENCE}
Let assume that monomolecular luminescent centers are excited by rectangular
pulse.
\begin{figure}[htbp]
  \centering
  \includegraphics[width=7.5cm]{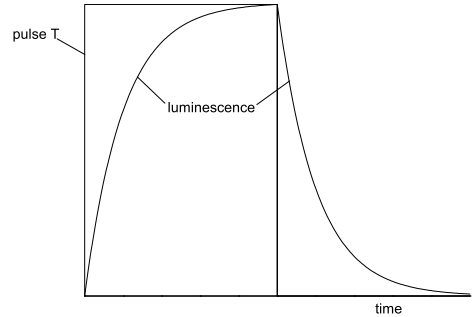}
  \caption{Time dependence of direct excitation luminescence excited by rectangular pulse}
  \label{fig:f1}
\end{figure}
If direct excitation is the only process then the luminescence rises
during the exciting pulse and begins to decay immediately after the end of
the pulse (Fig.~\ref{fig:f1}). For monomolecular centers we have
\begin{eqnarray}
\label{eq:dndt}
\frac{dn}{dt}&=&G-\alpha n \, ,
\end{eqnarray}
where $G$ is generation rate and $n$ is number of excited luminescent
centers at the time $t$. Resolving the equation we obtain
\begin{eqnarray}
\label{eq:n}
n&=&\frac{G}{\alpha} (1-e^{-\alpha t})+ n_{0} e^{-\alpha t} \, ,
\end{eqnarray}
where $n_{0}$ is number of the excited centers at the end of the
pulse. Taking into account that for monomolecular kinetic the
luminescence intensity is $I=\alpha n$ and that lifetime $\tau$
is $1/\alpha$ we have
\begin{eqnarray}
\label{eq:I}
I&=&G (1-e^{-t/\tau})+ I_{0} e^{-t/\tau} \, ,
\end{eqnarray}
where $I_{0}=\alpha n_{0}$ is the luminescence intensity at the end
of the pulse. If we assume that, at the start, the luminescent centers
were not excited than the equation
\begin{eqnarray}
\label{eq:Id}
I_d&=&G (1-e^{-t/\tau}) \, ,
\end{eqnarray}
describes luminescence intensity during the exciting pulse and equation
\begin{eqnarray}
\label{eq:Ia}
I_a&=&I_{0} (1-e^{-t/\tau}) \, ,
\end{eqnarray}
describes luminescence intensity after the end of the pulse. From
the equation we can get formula used to determination of luminescence
lifetime
\begin{eqnarray}
\label{eq:tau}
\tau &=&\frac{t}{\ln(I_{0}/I)}  \, .
\end{eqnarray}
Integrated light of direct excitation luminescence during the exciting
pulse, obtained from equation~(\ref{eq:Id}), has form
\begin{eqnarray}
\label{eq:Sd}
S_{d} &=&\int^{T}_{0} I_{d}dt=G T+G\tau e^{-T/\tau}-G\tau  \, .
\end{eqnarray}
After the end of the pulse the integrated light is described as
\begin{eqnarray}
\label{eq:Sa}
S_{a} &=&\int^{\infty}_{0} I_{a}dt=G\tau+G\tau e^{-T/\tau}  \, .
\end{eqnarray}
Total integrated light of direct excitation luminescence is
\begin{eqnarray}
\label{eq:S}
S &=&S_{d}+S_{a}=G T  \, .
\end{eqnarray}

\section{ENERGY TRANSFER LUMINESCENCE}
\subsection{Luminescence excited by short pulse}
In some cases, luminescence maximum appears after a time $t_m$
($t_{m0}$ for very short pulse), considerably longer than the pulse
duration $T$. Such behavior can be explained by assuming energy
transfer between two monomolecular centers. The number of the
transferring centers $dn_1$ which recombine during the time $dt$ is
\begin{eqnarray}
\label{eq:dn1}
dn_{1} &=&-\alpha_{1} n_{1} dt  \, .
\end{eqnarray}
The change in the excited emitting centers is described as
\begin{eqnarray}
\label{eq:dn2}
dn_{2} &=&dn_{2}'+dn_{2}''  \, ,
\end{eqnarray}
where $dn_{2}'= -dn_{1}$ and $dn_{2}''= -\alpha_{2}n_{2}dt$.
After mathematical treatments, the time dependence of the luminescence
is given as
\begin{eqnarray}
\label{eq:Ib}
I &=&-\frac{dn_{2}''}{dt}=\frac{n_{01}}{\tau_{2}-\tau_{1}}\Big(e^{-t/\tau_2}-e^{-t/\tau_1}\Big)  \, ,
\end{eqnarray}
where $n_{01}$ is the initial number of excited transferring centers
and $\tau_1$, $\tau_2$ are the lifetimes of the transferring and
emitting centers, respectively.
\begin{figure}[htbp]
  \centering
  \includegraphics[width=7.5cm]{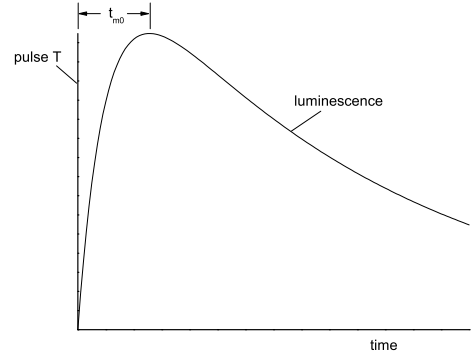}
  \caption{Time dependence of energy transfer luminescence excited by short pulse}
  \label{fig:f2}
\end{figure}
The relation (12) is shown in Fig.~\ref{fig:f2}. The curve in the
figure attains its maximum at the time
\begin{eqnarray}
\label{eq:tm0}
t_{m0} &=&\frac{\tau_1\tau_2}{\tau_2-\tau_1} \ln\frac{\tau_2}{\tau_1} \, ,
\end{eqnarray}
Both the last relations are in good agreement with experimental data
for electroluminescence of thin film cells based on 
ZnS:Mn~\cite{chimczak84,bertrandt95}.

\subsection{Time dependence of luminescence during rectangular exciting pulse}
\begin{figure}[htbp]
  \centering
  \includegraphics[width=7.5cm]{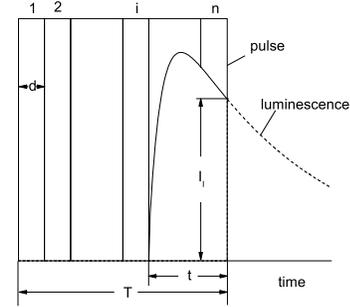}
  \caption{Luminescence produced at the end of exciting pulse T by short parts d of the pulse}
  \label{fig:f3}
\end{figure}

\begin{figure}[htbp]
  \centering
  \includegraphics[width=7.5cm]{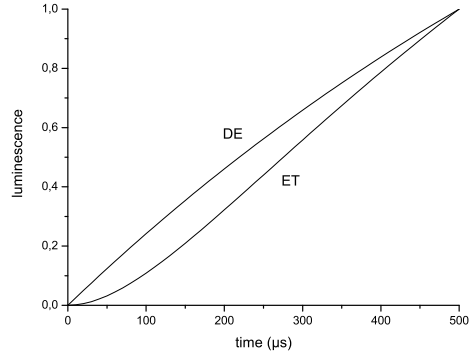}
  \caption{Time dependence of luminescence during exciting
           rectangular pulse of 500 $\mu s$.
           DE - direct excitation, ET - energy transfer}
  \label{fig:f4b}
\end{figure}

\begin{figure}[htbp]
  \centering
  \includegraphics[width=7.5cm]{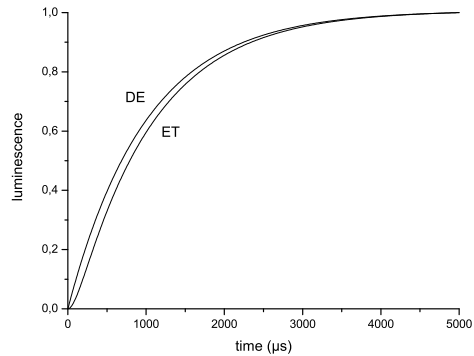}
  \caption{Time dependence of luminescence during exciting
           rectangular pulse of 5000 $\mu s$.
           DE - direct excitation, ET - energy transfer}
  \label{fig:f4c}
\end{figure}

Let us assume, to explain the time dependence, that each part $d$ of
the exciting pulse $T$ produces the elemental curve~(\ref{eq:Ib})
(Fig.~\ref{fig:f3})~\cite{chimczak88lum}.
After the short part $d$
of the exciting pulse $T$, initial number of excited transferring
centers is
\begin{eqnarray}
\label{eq:n01}
n_{01} &=&G d \, .
\end{eqnarray}
Intensity of the luminescence, at the end of the pulse $T$,
produced by the part $d_i$ is
\begin{eqnarray}
\label{eq:Ii}
I_{i} &=&\frac{G d}{\tau_{2}-\tau_{1}}
\Big(e^{-(n-i)d/\tau_{2}}-e^{-(n-i)d/\tau_{1}}\Big) \, .
\end{eqnarray}
At the end of the exciting pulse the luminescence intensity is
\begin{eqnarray}
\label{eq:Isum}
I &=&\sum_{i=1}^{n-1} I_{i} \, ,
\end{eqnarray}
In order to calculate eq.~(\ref{eq:Isum}) we will use
\begin{eqnarray}
\label{eq:toz}
d e^{-\frac{d}{\tau}}+d e^{-\frac{2 d}{\tau}}+\dots && \nonumber \\
+d e^{-\frac{(n-1)d}{\tau}}&=&d\frac{e^{-d/\tau}(e^{-(n-1)d/\tau}-1)}{e^{-d/\tau}-1} \, .
\end{eqnarray}
Taking into account that $T=nd$, we have
\begin{eqnarray}
\label{eq:lim}
\lim_{d\to 0}\frac{e^{-T/\tau}-e^{-d/\tau}+d e^{-d/\tau}/\tau}{-e^{-d/\tau}/\tau}
&=&\tau(1-e^{-\frac{T}{\tau}}) \, .
\end{eqnarray}
On insertion of eq.~(\ref{eq:lim}) into eq.~(\ref{eq:Isum}) the luminescence
intensity at the end of the exciting pulse $T$ is given by
\begin{eqnarray}
\label{eq:Ic}
I&=&\frac{G}{\tau_{2}-\tau_{1}}\Big(\tau_{2}(1-e^{-T/\tau_{2}})
-\tau_{1}(1-e^{-T/\tau_{1}})\Big) \, .
\end{eqnarray}
If the relation~(\ref{eq:Ic}) is assumed to be valid at any time
$t\leq T$, the time dependence of the luminescence, during the
exciting pulse $T$, is described as
\begin{eqnarray}
\label{eq:Ict}
I&=&\frac{G}{\tau_{2}-\tau_{1}}\Big(\tau_{2}(1-e^{-t/\tau_{2}})
-\tau_{1}(1-e^{-t/\tau_{1}})\Big) \, .
\end{eqnarray}
This function, contrary to direct excitation luminescence, has
got a point of inflection. This point appears at the same time
$t_{m0}$, at which the first elemental luminescence curve
(Fig.~\ref{fig:f3}) excited by the first part $d$ of the pulse $T$
attains its maximum. The luminescence strongly increases with time
until the time $t_{m0}$, when it tends to saturate. Equation (20)
very well describes the electroluminescence of ZnS:Mn thin
films~\cite{chimczak89}. 
\begin{figure}[htbp]
  \centering
  \includegraphics[width=7.5cm]{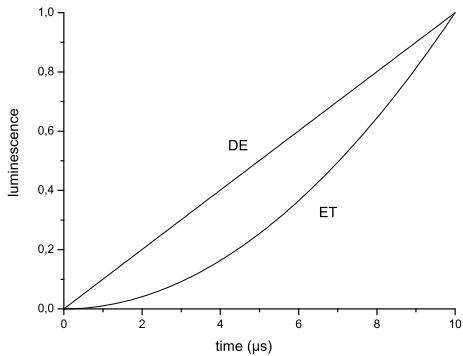}
  \caption{Time dependence of luminescence during exciting
           rectangular pulse of 10 $\mu s$. DE - direct
           excitation, ET - energy transfer}
  \label{fig:f4a}
\end{figure}
Figures~\ref{fig:f4a},~\ref{fig:f4b} and~\ref{fig:f4c}
show relative comparison of both, direct
excitation (DE) and energy transfer (ET), mechanisms. The curves
in these figures were plotted at $\tau_1= 100\mu s$ and
$\tau_2= 1000\mu s$. When the luminescence is excited by very short
pulse of $10\mu s$ (Fig.~\ref{fig:f4a}), then the luminescence at the
first stage of excitation is practically not seen. In the case,
as is seen in the figure, there is great difference between both
mechanisms.
For a longer width of the exciting pulse (Fig.~\ref{fig:f4b} and~\ref{fig:f4c}),
the difference in time dependence of both mechanisms disappears.
Also, the difference disappears when the lifetime, $\tau_1$, of
the transferring centre decreases. The small difference between
both mechanisms in the case of great width of the exciting pulse
or short lifetime of the transferring centre are probably the reason
why experimental dependences similar to that of Fig.~\ref{fig:f4c} are sometimes
faulty assumed to be due to direct excitation. Integrated light of
energy transfer luminescence during the rectangular exciting pulse,
obtained from equation~(\ref{eq:Ict}), has form
\begin{eqnarray}
\label{eq:Sd2}
S_{d}&=&G T-\frac{G\tau_{2}^{2}}{\tau_{2}-\tau_{1}}\big(1-e^{-\frac{T}{\tau_{2}}}\big)
+\frac{G\tau_{1}^{2}}{\tau_{2}-\tau_{1}}\big(1-e^{-\frac{T}{\tau_{1}}}\big) \, . \nonumber \\
\end{eqnarray}

\subsection{Time dependence of luminescence after the end of rectangular exciting pulse}

\begin{figure}[htbp]
  \centering
  \includegraphics[width=7.5cm]{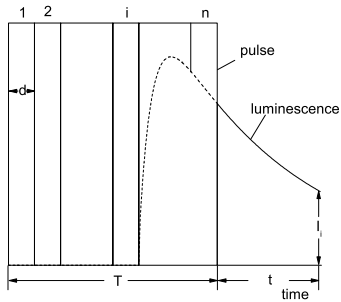}
  \caption{Luminescence produced at the time $t$ after exciting pulse $T$ by short parts d of the pulse}
  \label{fig:f5}
\end{figure}

Similar to that of Sect.~B, we assume that the exciting pulse consists
of $n$ short pulses $d$ (Fig.~\ref{fig:f5})~\cite{chimczak88zPhys}.
The luminescence produced by part $d_i$ of the pulse $T$ is
\begin{eqnarray}
\label{eq:Ii2}
I_{i} &=&\frac{G d}{\tau_{2}-\tau_{1}}
\Big(e^{-\frac{t+(n-i)d}{\tau_{2}}}-e^{-\frac{t+(n-i)d}{\tau_{1}}}\Big) \, .
\end{eqnarray}
Using the same procedure, we have got the relation describing the
time dependence of luminescence after the end of the exciting pulse,
for energy transfer between two monomolecular centers as
\begin{eqnarray}
\label{eq:I3}
I&=&\frac{G}{\tau_{2}-\tau_{1}}
\Big(\tau_{2}(1-e^{-\frac{T}{\tau_{2}}})e^{-\frac{t}{\tau_{2}}}
-\tau_{1}(1-e^{-\frac{T}{\tau_{1}}})e^{-\frac{t}{\tau_{1}}}\Big) \, . 
\nonumber \\
\end{eqnarray}
\begin{figure}[htbp]
  \centering
  \includegraphics[width=7.5cm]{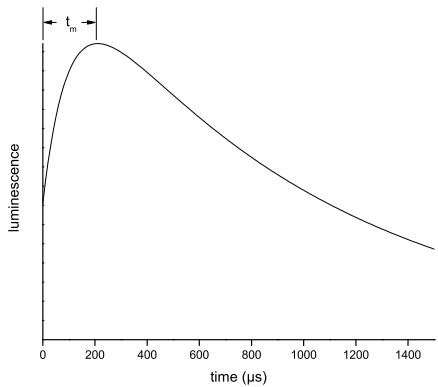}
  \caption{Time dependence of energy transfer luminescence after exciting rectangular pulse}
  \label{fig:f6}
\end{figure}
Figure~\ref{fig:f6} shows the time dependence described by the equation~(\ref{eq:I3}).
For short exciting pulses, the luminescence intensity at the maximum, after
the end of the exciting pulse, considerably exceeds the value of the
luminescence intensity at the end of the pulse. The difference is smaller
for longer pulse-duration and practically absent for very long pulses.
However, even in the last case, the luminescence attains its maximum,
after the end of the exciting pulse, at a time
\begin{eqnarray}
\label{eq:tm2}
t_{m} &=&\frac{\tau_{1}\tau_{2}}{\tau_{2}-\tau_{1}}
\ln{\frac{e^{-T/\tau_{1}}-1}{e^{-T/\tau_{2}}-1}} \, .
\end{eqnarray}
As is seen, the time $t_{m}$ depends on $T$. When $T$ is going to 0, $t_{m}$
is going to $t_{m0}$ and the luminescence curve is going to that described
by~(\ref{eq:Ib}). When $T$ is going to infinity, $t_{m}$ is going to 0
and the time dependence is going to that described by~(\ref{eq:Ia}).
Contrary to direct excitation, the luminescence
after the end of the exciting pulse~(\ref{eq:I3}) also
depends on $T$. The equation~(\ref{eq:I3}) explains why the first
experimental points, after the end of the exciting pulse, lie
below straight line when time dependence of luminescence is plotted
in semilogarithmic scale. From the~(\ref{eq:I3}), we obtain the
integrated light of energy transfer luminescence after the end of
the rectangular exciting pulse
\begin{eqnarray}
\label{eq:Sa2}
S_{a}&=&\frac{G\tau_{2}^{2}}{\tau_{2}-\tau_{1}}\big(1-e^{-\frac{T}{\tau_{2}}}\big)
-\frac{G\tau_{1}^{2}}{\tau_{2}-\tau_{1}}\big(1-e^{-\frac{T}{\tau_{1}}}\big) \, .
\end{eqnarray}
Total integrated light of energy transfer luminescence is
\begin{eqnarray}
\label{eq:totalS}
S&=&G T \, .
\end{eqnarray}

\section{Conclusion}
Time dependence of luminescence during and after a rectangular
exciting pulse was discussed. In the case of direct excitation
of luminescence, the intensity of the luminescence increases
during the exciting pulse and begins to decay immediately after
the end of the pulse. When there is energy transfer to luminescence
center from another center then the luminescence can increase in
intensity after the end of the excitation pulse before subsequently
decaying. During the exciting pulse the luminescence is practically
not seen at the first stage of excitation. The luminescence curve
has got a point of inflection at the time $t_{m0}$~(\ref{eq:tm0}).
The energy transfer luminescence attains its maximum after the end
of the exciting pulse at the time $t_{m}$~(\ref{eq:tm2}). In
the case of very short exciting pulses, there is a great difference
between both mechanisms. For a long width of exciting pulse or very
short lifetime of the transferring center, the difference disappears.
It is the reason why it is sometimes faulty assumed that, for longer
pulse lengths, the dominant excitation mechanism is direct excitation.
It is shown in the paper, that total integrated light is proportional
to the pulse length for direct excitation as well as energy transfer
luminescence. From the experiment, it exhibits that, for short exciting
pulses, the integrated light is not proportional to the excitation pulse
length~\cite{chimczak85,rigby88}. There is some excess in the region
of short pulses. The results obtained in the paper show that the
excess is not due to energy transfer mechanism.

\section*{ACKNOWLEDGMENTS}
This work was supported by Pozna\'n University of Technology Research 
Program 62-217/07 - BW.


\begin{thebibliography}{34}
\expandafter\ifx\csname natexlab\endcsname\relax\def\natexlab#1{#1}\fi
\expandafter\ifx\csname bibnamefont\endcsname\relax
  \def\bibnamefont#1{#1}\fi
\expandafter\ifx\csname bibfnamefont\endcsname\relax
  \def\bibfnamefont#1{#1}\fi
\expandafter\ifx\csname citenamefont\endcsname\relax
  \def\citenamefont#1{#1}\fi
\expandafter\ifx\csname url\endcsname\relax
  \def\url#1{\texttt{#1}}\fi
\expandafter\ifx\csname urlprefix\endcsname\relax\def\urlprefix{URL }\fi
\providecommand{\bibinfo}[2]{#2}
\providecommand{\eprint}[2][]{\url{#2}}

\bibitem[{\citenamefont{Chimczak et~al.}(1984)\citenamefont{Chimczak, Gordon,
  and Bertrandt-\.Zytkowiak}}]{chimczak84}
\bibinfo{author}{\bibfnamefont{E.}~\bibnamefont{Chimczak}},
  \bibinfo{author}{\bibfnamefont{W.~S.} \bibnamefont{Gordon}},
  \bibnamefont{and}
  \bibinfo{author}{\bibfnamefont{M.}~\bibnamefont{Bertrandt-\.Zytkowiak}},
  \bibinfo{journal}{Phys. Stat. Sol.~(a)} \textbf{\bibinfo{volume}{82}},
  \bibinfo{pages}{527} (\bibinfo{year}{1984}).

\bibitem[{\citenamefont{Bertrandt-\.Zytkowiak and
  Chimczak}(1995)}]{bertrandt95}
\bibinfo{author}{\bibfnamefont{M.}~\bibnamefont{Bertrandt-\.Zytkowiak}}
  \bibnamefont{and} \bibinfo{author}{\bibfnamefont{E.}~\bibnamefont{Chimczak}},
  \bibinfo{journal}{Thin Solid Films} \textbf{\bibinfo{volume}{256}},
  \bibinfo{pages}{136} (\bibinfo{year}{1995}).

\bibitem[{\citenamefont{Chimczak and
  Bertrandt-\.Zytkowiak}(1989{\natexlab{a}})}]{chimczak89}
\bibinfo{author}{\bibfnamefont{E.}~\bibnamefont{Chimczak}} \bibnamefont{and}
  \bibinfo{author}{\bibfnamefont{M.}~\bibnamefont{Bertrandt-\.Zytkowiak}},
  \bibinfo{journal}{Phys. Stat. Sol.~(a)} \textbf{\bibinfo{volume}{115}},
  \bibinfo{pages}{581} (\bibinfo{year}{1989}{\natexlab{a}}).

\bibitem[{\citenamefont{Chimczak and
  Bertrandt-\.Zytkowiak}(1989{\natexlab{b}})}]{chimczak89acta}
\bibinfo{author}{\bibfnamefont{E.}~\bibnamefont{Chimczak}} \bibnamefont{and}
  \bibinfo{author}{\bibfnamefont{M.}~\bibnamefont{Bertrandt-\.Zytkowiak}},
  \bibinfo{journal}{Acta Phys. Pol. A} \textbf{\bibinfo{volume}{77}},
  \bibinfo{pages}{399} (\bibinfo{year}{1989}{\natexlab{b}}).

\bibitem[{\citenamefont{Berdowski
  et~al.}(1985{\natexlab{a}})\citenamefont{Berdowski, Lammers, and
  Blasse}}]{berdowski85chem}
\bibinfo{author}{\bibfnamefont{P.~A.~M.} \bibnamefont{Berdowski}},
  \bibinfo{author}{\bibfnamefont{M.~J.~J.} \bibnamefont{Lammers}},
  \bibnamefont{and} \bibinfo{author}{\bibfnamefont{G.}~\bibnamefont{Blasse}},
  \bibinfo{journal}{J. Chem. Phys.} \textbf{\bibinfo{volume}{83}},
  \bibinfo{pages}{476} (\bibinfo{year}{1985}{\natexlab{a}}).

\bibitem[{\citenamefont{Wang et~al.}(2003)\citenamefont{Wang, Wu, Chen, and
  Huang}}]{wang03}
\bibinfo{author}{\bibfnamefont{Y.-J.} \bibnamefont{Wang}},
  \bibinfo{author}{\bibfnamefont{C.-X.} \bibnamefont{Wu}},
  \bibinfo{author}{\bibfnamefont{M.-Z.} \bibnamefont{Chen}}, \bibnamefont{and}
  \bibinfo{author}{\bibfnamefont{M.-C.} \bibnamefont{Huang}},
  \bibinfo{journal}{J. Appl. Phys.} \textbf{\bibinfo{volume}{93}},
  \bibinfo{pages}{9625} (\bibinfo{year}{2003}).

\bibitem[{\citenamefont{Kimpel et~al.}(1995)\citenamefont{Kimpel, Lobe, Schulz,
  and Zeitler}}]{kimpel95}
\bibinfo{author}{\bibfnamefont{B.~M.} \bibnamefont{Kimpel}},
  \bibinfo{author}{\bibfnamefont{K.}~\bibnamefont{Lobe}},
  \bibinfo{author}{\bibfnamefont{H.~J.} \bibnamefont{Schulz}},
  \bibnamefont{and} \bibinfo{author}{\bibfnamefont{E.}~\bibnamefont{Zeitler}},
  \bibinfo{journal}{Meas. Sci. Technol.} \textbf{\bibinfo{volume}{6}},
  \bibinfo{pages}{1383} (\bibinfo{year}{1995}).

\bibitem[{\citenamefont{Roura et~al.}(1994)\citenamefont{Roura, Costa, Sardin,
  Morante, and Bertran}}]{roura94}
\bibinfo{author}{\bibfnamefont{P.}~\bibnamefont{Roura}},
  \bibinfo{author}{\bibfnamefont{J.}~\bibnamefont{Costa}},
  \bibinfo{author}{\bibfnamefont{G.}~\bibnamefont{Sardin}},
  \bibinfo{author}{\bibfnamefont{J.~R.} \bibnamefont{Morante}},
  \bibnamefont{and} \bibinfo{author}{\bibfnamefont{E.}~\bibnamefont{Bertran}},
  \bibinfo{journal}{Phys. Rev. B} \textbf{\bibinfo{volume}{50}},
  \bibinfo{pages}{18124} (\bibinfo{year}{1994}).

\bibitem[{\citenamefont{Bhargava et~al.}(1994)\citenamefont{Bhargava,
  Gallagher, Hong, and Nurmikko}}]{bhargava94}
\bibinfo{author}{\bibfnamefont{R.~N.} \bibnamefont{Bhargava}},
  \bibinfo{author}{\bibfnamefont{D.}~\bibnamefont{Gallagher}},
  \bibinfo{author}{\bibfnamefont{X.}~\bibnamefont{Hong}}, \bibnamefont{and}
  \bibinfo{author}{\bibfnamefont{A.}~\bibnamefont{Nurmikko}},
  \bibinfo{journal}{Phys. Rev. Lett.} \textbf{\bibinfo{volume}{72}},
  \bibinfo{pages}{416} (\bibinfo{year}{1994}).

\bibitem[{\citenamefont{Krasnov and Hofstra}(2001)}]{krasnov01}
\bibinfo{author}{\bibfnamefont{A.~N.} \bibnamefont{Krasnov}} \bibnamefont{and}
  \bibinfo{author}{\bibfnamefont{P.~G.} \bibnamefont{Hofstra}},
  \bibinfo{journal}{Prog. Crystal Growth and Charact.}
  \textbf{\bibinfo{volume}{42}}, \bibinfo{pages}{65} (\bibinfo{year}{2001}).

\bibitem[{\citenamefont{John and Singh}(1996)}]{john96}
\bibinfo{author}{\bibfnamefont{G.~C.} \bibnamefont{John}} \bibnamefont{and}
  \bibinfo{author}{\bibfnamefont{V.~A.} \bibnamefont{Singh}},
  \bibinfo{journal}{Phys. Rev. B} \textbf{\bibinfo{volume}{54}},
  \bibinfo{pages}{4416} (\bibinfo{year}{1996}).

\bibitem[{\citenamefont{Barthou et~al.}(1994)\citenamefont{Barthou, Benoit,
  Benalloul, and Morell}}]{barthou94}
\bibinfo{author}{\bibfnamefont{C.}~\bibnamefont{Barthou}},
  \bibinfo{author}{\bibfnamefont{J.}~\bibnamefont{Benoit}},
  \bibinfo{author}{\bibfnamefont{P.}~\bibnamefont{Benalloul}},
  \bibnamefont{and} \bibinfo{author}{\bibfnamefont{A.}~\bibnamefont{Morell}},
  \bibinfo{journal}{J. Electrochem. Soc.} \textbf{\bibinfo{volume}{141}},
  \bibinfo{pages}{524} (\bibinfo{year}{1994}).

\bibitem[{\citenamefont{Zhang et~al.}(1995{\natexlab{a}})\citenamefont{Zhang,
  Sturge, Kash, van~der Gaag, Gozdz, Florez, and Harbison}}]{zhang95PRB}
\bibinfo{author}{\bibfnamefont{Y.}~\bibnamefont{Zhang}},
  \bibinfo{author}{\bibfnamefont{M.~D.} \bibnamefont{Sturge}},
  \bibinfo{author}{\bibfnamefont{K.}~\bibnamefont{Kash}},
  \bibinfo{author}{\bibfnamefont{B.~P.} \bibnamefont{van~der Gaag}},
  \bibinfo{author}{\bibfnamefont{A.~S.} \bibnamefont{Gozdz}},
  \bibinfo{author}{\bibfnamefont{L.~T.} \bibnamefont{Florez}},
  \bibnamefont{and} \bibinfo{author}{\bibfnamefont{J.~P.}
  \bibnamefont{Harbison}}, \bibinfo{journal}{Phys. Rev. B}
  \textbf{\bibinfo{volume}{51}}, \bibinfo{pages}{13303}
  (\bibinfo{year}{1995}{\natexlab{a}}).

\bibitem[{\citenamefont{Busse et~al.}(1976)\citenamefont{Busse, Gumlich,
  Meissner, and Theis}}]{busse76}
\bibinfo{author}{\bibfnamefont{W.}~\bibnamefont{Busse}},
  \bibinfo{author}{\bibfnamefont{H.~E.} \bibnamefont{Gumlich}},
  \bibinfo{author}{\bibfnamefont{B.}~\bibnamefont{Meissner}}, \bibnamefont{and}
  \bibinfo{author}{\bibfnamefont{D.}~\bibnamefont{Theis}}, \bibinfo{journal}{J.
  Lumin.} \textbf{\bibinfo{volume}{12-13}}, \bibinfo{pages}{693}
  (\bibinfo{year}{1976}).

\bibitem[{\citenamefont{Seo et~al.}(2003)\citenamefont{Seo, Kim, and
  Shin}}]{seo03}
\bibinfo{author}{\bibfnamefont{S.-Y.} \bibnamefont{Seo}},
  \bibinfo{author}{\bibfnamefont{M.-J.} \bibnamefont{Kim}}, \bibnamefont{and}
  \bibinfo{author}{\bibfnamefont{J.~H.} \bibnamefont{Shin}},
  \bibinfo{journal}{Appl. Phys. Lett.} \textbf{\bibinfo{volume}{83}},
  \bibinfo{pages}{2778} (\bibinfo{year}{2003}).

\bibitem[{\citenamefont{Weman et~al.}(1995)\citenamefont{Weman, Harris,
  Bergman, Miller, Yi, and Merz}}]{weman95}
\bibinfo{author}{\bibfnamefont{H.}~\bibnamefont{Weman}},
  \bibinfo{author}{\bibfnamefont{C.~J.} \bibnamefont{Harris}},
  \bibinfo{author}{\bibfnamefont{J.~P.} \bibnamefont{Bergman}},
  \bibinfo{author}{\bibfnamefont{M.~S.} \bibnamefont{Miller}},
  \bibinfo{author}{\bibfnamefont{J.~C.} \bibnamefont{Yi}}, \bibnamefont{and}
  \bibinfo{author}{\bibfnamefont{J.~L.} \bibnamefont{Merz}},
  \bibinfo{journal}{Superlatt. and Microstruct} \textbf{\bibinfo{volume}{17}},
  \bibinfo{pages}{61} (\bibinfo{year}{1995}).

\bibitem[{\citenamefont{Harukawa et~al.}(2000)\citenamefont{Harukawa, Murakami,
  Tamon, Ijuin, Ohmori, Abe, and Shigenari}}]{harukawa00}
\bibinfo{author}{\bibfnamefont{N.}~\bibnamefont{Harukawa}},
  \bibinfo{author}{\bibfnamefont{S.}~\bibnamefont{Murakami}},
  \bibinfo{author}{\bibfnamefont{S.}~\bibnamefont{Tamon}},
  \bibinfo{author}{\bibfnamefont{S.}~\bibnamefont{Ijuin}},
  \bibinfo{author}{\bibfnamefont{A.}~\bibnamefont{Ohmori}},
  \bibinfo{author}{\bibfnamefont{K.}~\bibnamefont{Abe}}, \bibnamefont{and}
  \bibinfo{author}{\bibfnamefont{T.}~\bibnamefont{Shigenari}},
  \bibinfo{journal}{J. Lumin.} \textbf{\bibinfo{volume}{87-89}},
  \bibinfo{pages}{1231} (\bibinfo{year}{2000}).

\bibitem[{\citenamefont{Joubert et~al.}(1987)\citenamefont{Joubert, Jacquier,
  Linar\`es, Chaminade, and Wanklyn}}]{joubert87}
\bibinfo{author}{\bibfnamefont{M.~F.} \bibnamefont{Joubert}},
  \bibinfo{author}{\bibfnamefont{B.}~\bibnamefont{Jacquier}},
  \bibinfo{author}{\bibfnamefont{C.}~\bibnamefont{Linar\`es}},
  \bibinfo{author}{\bibfnamefont{J.~P.} \bibnamefont{Chaminade}},
  \bibnamefont{and} \bibinfo{author}{\bibfnamefont{B.~M.}
  \bibnamefont{Wanklyn}}, \bibinfo{journal}{J. Lumin.}
  \textbf{\bibinfo{volume}{37}}, \bibinfo{pages}{239} (\bibinfo{year}{1987}).

\bibitem[{\citenamefont{de~Lucas et~al.}(1994)\citenamefont{de~Lucas, Rodrigez,
  and Moreno}}]{lucas94}
\bibinfo{author}{\bibfnamefont{M.~C.} \bibnamefont{de~Lucas}},
  \bibinfo{author}{\bibfnamefont{F.}~\bibnamefont{Rodrigez}}, \bibnamefont{and}
  \bibinfo{author}{\bibfnamefont{M.}~\bibnamefont{Moreno}},
  \bibinfo{journal}{Phys. Stat. Sol.~(b)} \textbf{\bibinfo{volume}{184}},
  \bibinfo{pages}{247} (\bibinfo{year}{1994}).

\bibitem[{\citenamefont{Wolfert et~al.}(1985)\citenamefont{Wolfert, Oomen, and
  Blasse}}]{wolfert85}
\bibinfo{author}{\bibfnamefont{A.}~\bibnamefont{Wolfert}},
  \bibinfo{author}{\bibfnamefont{E.~W. J.~L.} \bibnamefont{Oomen}},
  \bibnamefont{and} \bibinfo{author}{\bibfnamefont{G.}~\bibnamefont{Blasse}},
  \bibinfo{journal}{J. Sol. Stat. Chem} \textbf{\bibinfo{volume}{59}},
  \bibinfo{pages}{280} (\bibinfo{year}{1985}).

\bibitem[{\citenamefont{Berdowski
  et~al.}(1985{\natexlab{b}})\citenamefont{Berdowski, van Herk, and
  Blasse}}]{berdowski85}
\bibinfo{author}{\bibfnamefont{P.~A.~M.} \bibnamefont{Berdowski}},
  \bibinfo{author}{\bibfnamefont{J.}~\bibnamefont{van Herk}}, \bibnamefont{and}
  \bibinfo{author}{\bibfnamefont{G.}~\bibnamefont{Blasse}},
  \bibinfo{journal}{J. Lumin.} \textbf{\bibinfo{volume}{34}},
  \bibinfo{pages}{9} (\bibinfo{year}{1985}{\natexlab{b}}).

\bibitem[{\citenamefont{Suchocki et~al.}(1987)\citenamefont{Suchocki,
  Gilliland, Powell, Bowen, and Walling}}]{suchocki87}
\bibinfo{author}{\bibfnamefont{A.~B.} \bibnamefont{Suchocki}},
  \bibinfo{author}{\bibfnamefont{G.~D.} \bibnamefont{Gilliland}},
  \bibinfo{author}{\bibfnamefont{C.}~\bibnamefont{Powell}},
  \bibinfo{author}{\bibfnamefont{M.}~\bibnamefont{Bowen}}, \bibnamefont{and}
  \bibinfo{author}{\bibfnamefont{J.~C.} \bibnamefont{Walling}},
  \bibinfo{journal}{J. Lumin.} \textbf{\bibinfo{volume}{37}},
  \bibinfo{pages}{29} (\bibinfo{year}{1987}).

\bibitem[{\citenamefont{Leslie and Allen}(1981)}]{leslie81}
\bibinfo{author}{\bibfnamefont{T.~C.} \bibnamefont{Leslie}} \bibnamefont{and}
  \bibinfo{author}{\bibfnamefont{J.~W.} \bibnamefont{Allen}},
  \bibinfo{journal}{Phys. Stat. Sol.~(a)} \textbf{\bibinfo{volume}{65}},
  \bibinfo{pages}{545} (\bibinfo{year}{1981}).

\bibitem[{\citenamefont{Chimczak and Allen}(1985)}]{chimczak85}
\bibinfo{author}{\bibfnamefont{E.}~\bibnamefont{Chimczak}} \bibnamefont{and}
  \bibinfo{author}{\bibfnamefont{J.~W.} \bibnamefont{Allen}},
  \bibinfo{journal}{J. Phys. D:~Appl. Phys.} \textbf{\bibinfo{volume}{18}},
  \bibinfo{pages}{951} (\bibinfo{year}{1985}).

\bibitem[{\citenamefont{Zhang et~al.}(1995{\natexlab{b}})\citenamefont{Zhang,
  Sturge, Kash, van~der Gaag, Gozdz, Florez, and Harbison}}]{zhang95}
\bibinfo{author}{\bibfnamefont{Y.}~\bibnamefont{Zhang}},
  \bibinfo{author}{\bibfnamefont{M.~D.} \bibnamefont{Sturge}},
  \bibinfo{author}{\bibfnamefont{K.}~\bibnamefont{Kash}},
  \bibinfo{author}{\bibfnamefont{B.~P.} \bibnamefont{van~der Gaag}},
  \bibinfo{author}{\bibfnamefont{A.~S.} \bibnamefont{Gozdz}},
  \bibinfo{author}{\bibfnamefont{L.~T.} \bibnamefont{Florez}},
  \bibnamefont{and} \bibinfo{author}{\bibfnamefont{J.~P.}
  \bibnamefont{Harbison}}, \bibinfo{journal}{Superlatt. and Microstruct.}
  \textbf{\bibinfo{volume}{17}}, \bibinfo{pages}{201}
  (\bibinfo{year}{1995}{\natexlab{b}}).

\bibitem[{\citenamefont{Bergman and Monemar}(1987)}]{bergman87}
\bibinfo{author}{\bibfnamefont{P.}~\bibnamefont{Bergman}} \bibnamefont{and}
  \bibinfo{author}{\bibfnamefont{B.}~\bibnamefont{Monemar}},
  \bibinfo{journal}{J. Lumin.} \textbf{\bibinfo{volume}{38}},
  \bibinfo{pages}{87} (\bibinfo{year}{1987}).

\bibitem[{\citenamefont{Lee et~al.}(1992)\citenamefont{Lee, Park, Chung, and
  Chang}}]{lee92}
\bibinfo{author}{\bibfnamefont{C.~D.} \bibnamefont{Lee}},
  \bibinfo{author}{\bibfnamefont{H.~L.} \bibnamefont{Park}},
  \bibinfo{author}{\bibfnamefont{C.~H.} \bibnamefont{Chung}}, \bibnamefont{and}
  \bibinfo{author}{\bibfnamefont{S.~K.} \bibnamefont{Chang}},
  \bibinfo{journal}{Phys. Rev. B} \textbf{\bibinfo{volume}{45}},
  \bibinfo{pages}{4491} (\bibinfo{year}{1992}).

\bibitem[{\citenamefont{Ramirez et~al.}(2005)\citenamefont{Ramirez, Jaque,
  Baus\'a, Mart\'in, Lahoz, Cavalli, Speghini, and Bettinelli}}]{ramirez05}
\bibinfo{author}{\bibfnamefont{M.~O.} \bibnamefont{Ramirez}},
  \bibinfo{author}{\bibfnamefont{D.}~\bibnamefont{Jaque}},
  \bibinfo{author}{\bibfnamefont{L.~E.} \bibnamefont{Baus\'a}},
  \bibinfo{author}{\bibfnamefont{I.~R.} \bibnamefont{Mart\'in}},
  \bibinfo{author}{\bibfnamefont{F.}~\bibnamefont{Lahoz}},
  \bibinfo{author}{\bibfnamefont{E.}~\bibnamefont{Cavalli}},
  \bibinfo{author}{\bibfnamefont{A.}~\bibnamefont{Speghini}}, \bibnamefont{and}
  \bibinfo{author}{\bibfnamefont{M.}~\bibnamefont{Bettinelli}},
  \bibinfo{journal}{J. Appl. Phys.} \textbf{\bibinfo{volume}{97}},
  \bibinfo{pages}{093510} (\bibinfo{year}{2005}).

\bibitem[{\citenamefont{Aizawa et~al.}(2006)\citenamefont{Aizawa, Sekiguchi,
  Katsumata, Komuro, and Morikawa}}]{aizawa06}
\bibinfo{author}{\bibfnamefont{H.}~\bibnamefont{Aizawa}},
  \bibinfo{author}{\bibfnamefont{M.}~\bibnamefont{Sekiguchi}},
  \bibinfo{author}{\bibfnamefont{T.}~\bibnamefont{Katsumata}},
  \bibinfo{author}{\bibfnamefont{S.}~\bibnamefont{Komuro}}, \bibnamefont{and}
  \bibinfo{author}{\bibfnamefont{T.}~\bibnamefont{Morikawa}},
  \bibinfo{journal}{Rev. Sci. Instr.} \textbf{\bibinfo{volume}{77}},
  \bibinfo{pages}{044902} (\bibinfo{year}{2006}).

\bibitem[{\citenamefont{Kapoor et~al.}(2000)\citenamefont{Kapoor, Singh, and
  Johri}}]{kapoor00}
\bibinfo{author}{\bibfnamefont{M.}~\bibnamefont{Kapoor}},
  \bibinfo{author}{\bibfnamefont{V.~A.} \bibnamefont{Singh}}, \bibnamefont{and}
  \bibinfo{author}{\bibfnamefont{G.~K.} \bibnamefont{Johri}},
  \bibinfo{journal}{Phys. Rev. B} \textbf{\bibinfo{volume}{61}},
  \bibinfo{pages}{1941} (\bibinfo{year}{2000}).

\bibitem[{\citenamefont{Thilsing-Hansen
  et~al.}(2005)\citenamefont{Thilsing-Hansen, Neves-Petersen, Petersen,
  Neuendorf, Al-Shamery, and Rubahn}}]{thilsing05}
\bibinfo{author}{\bibfnamefont{K.}~\bibnamefont{Thilsing-Hansen}},
  \bibinfo{author}{\bibfnamefont{M.~T.} \bibnamefont{Neves-Petersen}},
  \bibinfo{author}{\bibfnamefont{S.~B.} \bibnamefont{Petersen}},
  \bibinfo{author}{\bibfnamefont{R.}~\bibnamefont{Neuendorf}},
  \bibinfo{author}{\bibfnamefont{K.}~\bibnamefont{Al-Shamery}},
  \bibnamefont{and} \bibinfo{author}{\bibfnamefont{H.~G.}
  \bibnamefont{Rubahn}}, \bibinfo{journal}{Phys. Rev. B}
  \textbf{\bibinfo{volume}{72}}, \bibinfo{pages}{115213}
  (\bibinfo{year}{2005}).

\bibitem[{\citenamefont{Chimczak}(1988{\natexlab{a}})}]{chimczak88lum}
\bibinfo{author}{\bibfnamefont{E.}~\bibnamefont{Chimczak}},
  \bibinfo{journal}{J. Lumin.} \textbf{\bibinfo{volume}{39}},
  \bibinfo{pages}{247} (\bibinfo{year}{1988}{\natexlab{a}}).

\bibitem[{\citenamefont{Chimczak}(1988{\natexlab{b}})}]{chimczak88zPhys}
\bibinfo{author}{\bibfnamefont{E.}~\bibnamefont{Chimczak}},
  \bibinfo{journal}{Z. Phys. B - Condensed Matter}
  \textbf{\bibinfo{volume}{72}}, \bibinfo{pages}{211}
  (\bibinfo{year}{1988}{\natexlab{b}}).

\bibitem[{\citenamefont{Rigby and Allen}(1988)}]{rigby88}
\bibinfo{author}{\bibfnamefont{N.~E.} \bibnamefont{Rigby}} \bibnamefont{and}
  \bibinfo{author}{\bibfnamefont{J.~W.} \bibnamefont{Allen}},
  \bibinfo{journal}{J. Lumin.} \textbf{\bibinfo{volume}{42}},
  \bibinfo{pages}{143} (\bibinfo{year}{1988}).

\end{thebibliography}
\end{document}